\documentclass[prd,twocolumn,superscriptaddress,showpacs,%
preprintnumbers,nofootinbib,amsmath]{revtex4}

\usepackage{bm}

\begin{document}

\preprint{CECS-PHY-05-09}
\preprint{gr-qc/0506004}

\title{Conformally Flat Noncircular Spacetimes}

\author{Eloy Ay\'on--Beato}\email{ayon-at-cecs.cl}
\affiliation{Centro~de~Estudios~Cient\'{\i}ficos~(CECS),%
~Casilla~1469,~Valdivia,~Chile.}
\affiliation{Departamento~de~F\'{\i}sica,%
~Centro~de~Investigaci\'on~y~de~Estudios~Avanzados~del~IPN,\\
~Apdo.~Postal~14--740,~07000,~M\'exico~D.F.,~M\'exico.}
\author{Cuauhtemoc Campuzano}\email{cuauhtemoc.campuzano-at-ucv.cl}
\affiliation{Instituto de F\'{\i}sica, Pontificia Universidad
Cat\'olica de Valpara\'{\i}so, Casilla 4950, Valpara\'{\i}so.}
\affiliation{Departamento~de~F\'{\i}sica,%
~Centro~de~Investigaci\'on~y~de~Estudios~Avanzados~del~IPN,\\
~Apdo.~Postal~14--740,~07000,~M\'exico~D.F.,~M\'exico.}
\author{Alberto Garc\'\i a}\email{aagarcia-at-fis.cinvestav.mx}
\affiliation{Departamento~de~F\'{\i}sica,%
~Centro~de~Investigaci\'on~y~de~Estudios~Avanzados~del~IPN,\\
~Apdo.~Postal~14--740,~07000,~M\'exico~D.F.,~M\'exico.}

\date{\today}

\begin{abstract}
The general metric for conformally flat stationary cyclic symmetric
noncircular spacetimes is explicitly given. In spite of the
complexity introduced by the inclusion of noncircular contributions,
the related metric is derived via the full integration of the
conformal flatness constraints. It is also shown that the conditions
for the existence of a rotation axis (axisymmetry) are the same ones
which restrict the above class of spacetimes to be static. As a
consequence, a known theorem by Collinson is just part of a more
general result: any conformally flat stationary cyclic symmetric
spacetime, even a noncircular one, is additionally axisymmetric if
and only if it is also static. Since recent astrophysical
motivations point in the direction of considering noncircular
configurations to describe magnetized neutron stars, the above
results seem to be relevant in this context.
\end{abstract}


\maketitle

\section{\label{sec:intro}Introduction}

One of the challenges in general relativity is the search for
interior configurations describing isolated rotating bodies
supporting their corresponding exterior gravitational fields.
Usually, the description of these rotating configurations is done
by means of circular spacetimes, i.e., stationary axisymmetric
spacetimes where the metric, in addition to be time and
rotation-angle independent, possesses also as isometries the
inversions of time and of the rotation angle. It is worthwhile to
mention that almost all the stationary axisymmetric configurations
reported in the literature belong to this class.

Nevertheless, there is no room for the circular idealization when
considering rotating neutron stars~\cite{Gourgoulhon:1993}
surrounded by strong toroidal magnetic fields ranging $\sim
10^{16}$ to $10^{17}$ G, see also~\cite{Ioka:2003dd,Ioka:2003nh}
and references therein. The circularity condition is a very severe
restriction, which fails to hold in spacetimes allowing the
existence of toroidal magnetic fields and meridional
flows~\cite{Ioka:2003nh}. Thus, to deal with such astrophysical
configurations one has to abandon the fulfillment of the
circularity condition and consider in consequence the wider
noncircular class of spacetimes.

Besides the above astrophysically motivated reasons to study
noncircular configurations, there are also purely theoretical ones.
As soon as Schwarzschild published his exterior spherically
symmetric static solution, he was able to determine its interior
solution modeled trough a perfect fluid with homogeneous density.
Later, on the light of the Petrov classification, it was established
that the Schwarzschild solution belongs to Petrov type--D, while the
interior Schwarzschild solution falls in the conformally flat
family. In 1973 Kerr reported his famous stationary axisymmetric
gravitational field corresponding to a field created by a rotating
body; this solution also belongs to Petrov type--D. The search for
the interior solution to the exterior Kerr metric began since that
time. Collinson established that a conformally flat stationary
axisymmetric spacetime is necessarily static \cite{Collinson:1976}.
Some stationary axisymmetric Petrov--D metrics coupled to perfect
fluid distributions have been reported in the literature, but non of
them allows for the matching with the Kerr metric. Recently, the
results of Ref.~\cite{Vera:2003cn} indicate that the matching
between an interior noncircular spacetime with an exterior circular
one is at least technically possible. This fact opens the
possibility of searching for interior solutions within the
noncircular class.

Moreover, the general metric for conformally flat stationary cyclic
symmetric circular metric has been reported recently
\cite{Garcia:2002gj}, see on this respect also
\cite{Barnes:2003gz,Garcia:2003is}. For this metric, being cyclic
but not axisymmetric, the circularity theorem \cite{Kundt:1966} does
not hold because of the lack of a rotation axis. The next step in
complexity, which is the main goal of the present work, is to
determine the metric for conformally flat stationary cyclic
symmetric noncircular spacetimes. In particular, from this more
general result one is able to derive the particular circular branch,
and if one requires the existence of an axis of symmetry one
recovers the staticity property of the considered class of metrics,
thus the Collinson theorem is just included within a more general
result.

In the next section the mathematical preliminaries needed in order
to study the spacetimes under consideration are introduced.
Specifically, the physical and geometrical details behind the
concepts of stationarity, cyclic symmetry, axisymmetry, circularity,
and staticity are clearly stated in order to make the work
self-contained. In Sec.~\ref{sec:CTh} the conformal flatness
constraints, consisting in the vanishing of the complex Weyl
components, are fully integrated for any stationary cyclic symmetric
spacetime including in general noncircular contributions.
Sec.~\ref{sec:staticity} is devoted to revise the conditions
guarantying staticity on the obtained spacetimes, and in
Sec.~\ref{sec:axisymmetry} the locus of the axis of symmetry, for
the configurations allowing one, is found. It is concluded that both
physical situations occur for the same configuration. Some
conclusions are given in Sec.~\ref{sec:conclu}. In appendix A the
explicit form of the complex components of the Weyl tensor for a
general, not necessarily circular, stationary cyclic symmetric
spacetime are given. In appendix B, the problem is addressed for a
singular case ($a=b$) not included in the generic treatment.

\section{\label{sec:SCSS}Stationary Cyclic Symmetric Spacetimes}

In this section we characterize stationary cyclic symmetric
spacetimes, see for example Ref.~\cite{Heusler:1996} for the
original definitions. A spacetime is \emph{stationary} if it admits
an asymptotically timelike Killing field $\bm{k}$. A spacetime is
called \emph{cyclic symmetric} if it is invariant under the action
of the one--parameter group $SO(2)$, it is assumed that the
corresponding Killing field $\bm{m}$ with closed integral curves is
spacelike. A cyclic symmetric spacetime is named \emph{axisymmetric}
if the fixed point set of the $SO(2)$ action, i.e., the
\emph{rotation axis}, is nonempty. A spacetime is called
\emph{stationary cyclic symmetric (axisymmetric)} if it is both
stationary and cyclic symmetric (axisymmetric) and if the Killing
fields $\bm{k}$ and $\bm{m}$ commute.

A stationary cyclic symmetric (axisymmetric) spacetime is said to
be \emph{circular} if the $2$-surfaces orthogonal to the Killing
fields $\bm{k}$ and $\bm{m}$ are integrable. This is equivalent to
satisfy the Frobenius integrability conditions
\begin{eqnarray}
\bm{m}\wedge\bm{k}\wedge\bm{dk}&=&0,\nonumber\\
\bm{k}\wedge\bm{m}\wedge\bm{dm}&=&0.\label{eq:circularity}
\end{eqnarray}
The circularity property means that locally the gravitational field
is not only independent of time and the rotation angle, but, it is
also invariant under the simultaneous inversion of time and the
angle. Almost all the literature related to stationary cyclic
symmetric (axisymmetric) spacetimes concerns only with the circular
case. This is due in part for simplicity, since in this case it is
possible to use the Lewis-Papapetrou ansatz for the metric.

As a last definition, a stationary spacetime is said to be
\emph{static} if the Killing field $\bm{k}$ is hypersurface
orthogonal. This occurs if and only if it satisfies
\begin{equation}\label{eq:staticity}
\bm{k}\wedge\bm{dk}=0,
\end{equation}
and it is equivalent to demand that locally the gravitational
field is not only time-independent but it is also invariant under
time-reversal.

In this work we are interested in noncircular spacetimes, i.e,
general stationary cyclic symmetric spacetimes not necessarily
restricted to satisfy the Frobenius integrability conditions
(\ref{eq:circularity}). The metric of such spacetimes can be written
as
\begin{eqnarray}
\bm{g}&=&\mathrm{e}^{-2Q}\biggl(-\frac{1}{a+b}[\bm{d\tau}+a\,\bm{d\sigma}
+\mathrm{Im}(M\bm{dz})]\nonumber\\
&&\qquad\qquad\quad{}\times[\bm{d\tau}-b\,\bm{d\sigma}
-\mathrm{Im}(N\bm{dz})]\nonumber\\
&&\qquad\quad{}+\mathrm{e}^{-2P}\bm{dz}\bm{d}\bar{\bm{z}}\biggr),
\label{eq:metric}
\end{eqnarray}
where $a$, $b$, $P$, and $Q$ are real functions and $M$ and $N$
are complex ones. Here the bar means complex conjugation, and
$\mathrm{Im}$ (respectively $\mathrm{Re}$) denotes the imaginary
(respectively real) part of a complex quantity. All the functions
depend on the coordinates $z$ and $\bar{z}$ only, since the
Killing fields realizing the stationary and cyclic isometries are
$\bm{k}=\bm{\partial_\tau}$ and $\bm{m}=\bm{\partial_\sigma}$. The
above metric has eight independent real functions, hence,
diffeomorphism invariance allows to make two other gauge
elections. This metric (\ref{eq:metric}) is invariant under the
rescaling $z\mapsto\int{g(z)^{-1}\mathrm{d}z}$ together with the
redefinitions $P\mapsto{P-\ln\sqrt{g(z)\bar{g}(\bar{z})}}$,
$M\mapsto{Mg(z)}$, and $N\mapsto{Ng(z)}$. We shall fix the gauge
just after exploiting this special symmetry in our calculations.

The noncircularity of the above metric can be realized from the
fact that the following quantities are not necessarily zero
\begin{eqnarray}
\nonumber
*(\bm{m}\wedge\bm{k}\wedge\bm{dk})&=&\frac{\mathrm{e}^{2(P-Q)}}{2(a+b)}
\mathrm{Re}\left(\frac{\partial(M-N)}{\partial\bar{z}}\right.\\
&&\qquad\qquad\quad{}-\left.\frac{M+N}{a+b}
\frac{\partial(a-b)}{\partial\bar{z}}\right),\quad~\\
\nonumber
*(\bm{k}\wedge\bm{m}\wedge\bm{dm})&=&\frac{\mathrm{e}^{2(P-Q)}}{2(a+b)}
\mathrm{Re}\left(a\frac{\partial{M}}{\partial\bar{z}}
+b\frac{\partial{N}}{\partial\bar{z}}\right.\\
&&\qquad\qquad\quad{}-\left.\frac{M+N}{a+b}
\frac{\partial(ab)}{\partial\bar{z}}\right),
\end{eqnarray}
where the star stands for the Hodge dual.

In order to evaluate the Weyl tensor it is more convenient to use
the Newman-Penrose formalism. One starts writing the metric as
\begin{equation}\label{eq:metric_t}
\bm{g}=2\,\bm{e}^1%\otimes
\bm{e}^2-2\,\bm{e}^3%\otimes
\bm{e}^4,
\end{equation}
using a complex null tetrad, which in the present case is chosen
as
\begin{subequations}\label{eq:terad}
\begin{eqnarray}
\bm{e}^1&=&\frac{1}{\sqrt{2}}\mathrm{e}^{-Q-P}\bm{dz},\\
\bm{e}^2&=&\frac{1}{\sqrt{2}}\mathrm{e}^{-Q-P}\bm{d}\bar{\bm{z}},\\
\bm{e}^3&=&\frac{1}{\sqrt{2}}\frac{\mathrm{e}^{-Q}}{\sqrt{a+b}}
\left(\bm{d\tau}-b\,\bm{d\sigma}
-\frac{N\bm{dz}-\bar{N}\bm{d}\bar{\bm{z}})}{2i}\right),\\
\bm{e}^4&=&\frac{1}{\sqrt{2}}\frac{\mathrm{e}^{-Q}}{\sqrt{a+b}}
\left(\bm{d\tau}+a\,\bm{d\sigma}
+\frac{M\bm{dz}-\bar{M}\bm{d}\bar{\bm{z}}}{2i}\right).\quad~
\end{eqnarray}
\end{subequations}
The related Weyl complex components are given in the Appendix
\ref{app:Weyl}.

\section{\label{sec:CTh}Solving the Conformal Flatness Constraints}

In order to find the general class of conformally flat stationary
cyclic symmetric spacetimes we demand the vanishing of all the
complex components of the Weyl tensor, i.e.,
$\Psi_0=\Psi_4=\Psi_2=\Psi_1=\Psi_3=0$, see
Appendix~\ref{app:Weyl}.

The complex components $\Psi_0$ and $\Psi_4$ are the same than in
the circular case ($M=0=N$) studied in
Refs.~\cite{Collinson:1976,Garcia:2002gj}. Hence, initially, we
shall apply the same strategy of these references. First, the
vanishing of the following combinations
\begin{eqnarray}
\Psi_0-\bar{\Psi}_4&=&2(a+b)\mathrm{e}^{2Q}\nonumber\\
&&{}\times\frac{\partial}{\partial{z}}
\left(\frac{\mathrm{e}^{2P}}{(a+b)^2}
\frac{\partial(a-b)}{\partial{z}}\right)=0, \label{eq:0-4}\\
\Psi_0\frac{\partial{b}}{\partial{z}}
+\bar{\Psi}_4\frac{\partial{a}}{\partial{z}}
&=&2(a+b)\mathrm{e}^{2(Q-P)}\nonumber\\
&&{}\times\frac{\partial}{\partial{z}}
\left(\frac{\mathrm{e}^{4P}}{(a+b)^2}
\frac{\partial{a}}{\partial{z}}
\frac{\partial{b}}{\partial{z}}\right)=0, \label{eq:0b-4a}
\end{eqnarray}
give rise to the following first order conditions
\begin{eqnarray}
\frac{\partial{a}}{\partial{z}}-\frac{\partial{b}}{\partial{z}}
&=&\bar{g}(\bar{z})(a+b)^2\,e^{-2P}, \label{eq:az-bz}\\
\frac{\partial{a}}{\partial{z}}\frac{\partial{b}}{\partial{z}}
&=&\bar{h}(\bar{z})(a+b)^2\,e^{-4P}, \label{eq:az*bz}
\end{eqnarray}
where $g$ and $h$ are integration functions.

Secondly, since the functions $a$, $b$, and $P$ are real
Eq.~(\ref{eq:az-bz}) implies
\begin{equation}\label{eq:(a-b)}
g(z)\frac{\partial}{\partial{z}}(a-b)
=\bar{g}(\bar{z})\frac{\partial}{\partial\bar{z}}(a-b).
\end{equation}
Rescaling the complex coordinate by,
\begin{equation}\label{eq:rescal}
\textstyle
(\tau,\sigma,z,\bar{z})\mapsto
\left(\tau,\sigma,\int{g(z)^{-1}\mathrm{d}z},
\int{\bar{g}(\bar{z})^{-1}\mathrm{d}\bar{z}}\right),
\end{equation}
the above equation allows to conclude that in terms of the new
complex coordinate: $a-b=F(z+\bar{z})$.

Using the rescaling (\ref{eq:rescal}), together with the
redefinitions $P\mapsto{P-\ln\sqrt{g(z)\bar{g}(\bar{z})}}$ and
$\bar{h}(\bar{z})\mapsto\bar{h}(\bar{z})/\bar{g}(\bar{z})^2$, in
Eqs.~(\ref{eq:az-bz}) and (\ref{eq:az*bz}) is equivalent to put
$g(z)=1$ without loosing generality. This is due to the fact that
metric (\ref{eq:metric}), as it was previously anticipated, is
invariant under such changes if we consider also the redefinitions
$M\mapsto{Mg(z)}$ and $N\mapsto{Ng(z)}$.

Combining the fact that $a-b=F(z+\bar{z})$ with the imaginary part
of the component $\Psi_2$,
\begin{equation}\label{eq:ImPsi2}
\mathrm{Im}(\Psi_2)=\frac{\mathrm{e}^{2(Q+P)}}{i(a+b)^2}
\left(\frac{\partial{a}}{\partial{z}}\frac{\partial{b}}{\partial\bar{z}}
-\frac{\partial{a}}{\partial\bar{z}}\frac{\partial{b}}{\partial{z}}\right)=0,
\end{equation}
we conclude that if $F\neq\mathrm{const.}$ then $a=a(z+\bar{z})$,
$b=b(z+\bar{z})$, and consequently $P=P(z+\bar{z})$ by virtue of
Eq.~(\ref{eq:az-bz}).

In what follows we analyze the generic case
$\mathrm{d}F/\mathrm{d}x\neq0$, leaving the study of the special
case $\mathrm{d}F/\mathrm{d}x=0$ for the Appendix \ref{app:a=b}.

With regard to the other integration function appearing in
Eq.~(\ref{eq:az*bz}), since the left hand side of this equation is
real then
$\bar{h}(\bar{z})=h(z)=\mathrm{const.}\equiv\epsilon{k}^2$, where
$\epsilon\equiv\pm1$ just encodes the sign of the constant (see
Ref.~\cite{Garcia:2002gj} for the transcendence of this sign).

Using the real and imaginary parts of $z$ as coordinates,
$z=x+i\,y$, Eqs.~(\ref{eq:az-bz}) and (\ref{eq:az*bz}) are now
expressed by
\begin{eqnarray}
\frac{\mathrm{d}a}{\mathrm{d}x}-\frac{\mathrm{d}b}{\mathrm{d}x}
&=&(a+b)^2\,\mathrm{e}^{-2P}, \label{eq:ax-bx}\\
\frac{\mathrm{d}a}{\mathrm{d}x}\frac{\mathrm{d}b}{\mathrm{d}x}
&=&\epsilon{k}^2(a+b)^2\,\mathrm{e}^{-4P}. \label{eq:ax*bx}
\end{eqnarray}

We are now ready to fix the gauge. We choose that the redefined
complex functions $M$ and $N$, after the rescaling
(\ref{eq:rescal}), are real functions. That is, our gauge
elections are $\bar{M}=M$ and $\bar{N}=N$. It is important to
emphasize that until is correct to make such election from the
beginning, it is useless since we loose the scaling freedom in the
metric which allows to fix one of the above integration functions.
In terms of the real coordinates $x$ and $y$ the metric is written
in this gauge as
\begin{eqnarray}
\bm{g}&=&\mathrm{e}^{-2Q}\biggl(-\frac{1}{a+b}(\bm{d\tau}+a\,\bm{d\sigma}
+M\bm{dy})\nonumber\\
&&\qquad\qquad\quad{}\times(\bm{d\tau}-b\,\bm{d\sigma}
-N\bm{dy})\nonumber\\
&&\qquad\quad{}+\mathrm{e}^{-2P}(\bm{dx}^2+\bm{dy}^2)\biggr),
\label{eq:metricy}
\end{eqnarray}
where now we have six real functions, $M$, $N$, and $Q$ depending
on $x$ and $y$, and $a$, $b$, and $P$ depending just on $x$. In
the above coordinates the four gauge elections are
$g_{\tau{x}}=g_{\sigma{x}}=g_{yx}=0$ and $g_{xx}-g_{yy}$ is
proportional to the remaining noncircular components.

Let us infer some consequences concerning the noncircular
components of the metric. Using that $a=a(x)$, $b=b(x)$, and
$P=P(x)$ the real part of the following combination is written as
\begin{equation}\label{eq:Re(3-1)}
\mathrm{Re}(\Psi_3-\Psi_1)=\frac{\sqrt{a+b}\,\mathrm{e}^{2Q+3P}}{8}
\frac{\partial^2}{\partial{x}\partial{y}}
\left(\frac{M+N}{a+b}\right)=0,
\end{equation}
which implies that the following function is separable in $x$ and
$y$, i.e.,
\begin{equation}\label{eq:sep1}
\frac{M+N}{a+b}=F_1(x)+F_2(y),
\end{equation}
where $F_1$ and $F_2$ are undetermined functions. Isolating $N$
from the above expression and inserting it in the real part of
$\Psi_1$ we obtain
\begin{equation}\label{eq:Re(3)}
\mathrm{Re}(\Psi_1)=-\frac{\mathrm{e}^{2Q+3P}}{8\sqrt{a+b}}
\frac{\partial^2}{\partial{x}\partial{y}}
\left[M-(F_1+F_2)a\right]=0,
\end{equation}
and hence
\begin{equation}\label{eq:M}
M(x,y)=[F_1(x)+F_2(y)]a(x)+F_3(x)+F_4(y),
\end{equation}
where $F_3$ and $F_4$ is another pair of undetermined functions.
Using Eq.~(\ref{eq:sep1}), $N$ is given by
\begin{equation}\label{eq:N}
N(x,y)=[F_1(x)+F_2(y)]b(x)-F_3(x)-F_4(y).
\end{equation}

The dependence on the coordinate $y$ of the functions $M$ and $N$
(and of the whole problem) is encoded in the functions $F_2$ and
$F_4$. However, such functions can be eliminated by shifting
appropriately the Killing coordinates, i.e., the coordinate change
\begin{equation}\label{eq:shift}
\textstyle
(\tau,\sigma,x,y)\mapsto\left(\tau+\int{F_4(y)\mathrm{d}y},
\sigma+\int{F_2(y)\mathrm{d}y},x,y\right),
\end{equation}
is equivalent to put $F_2=0=F_4$. We would like to emphasize that
due to the same argument the functions $F_1$ and $F_3$ must be
determined up to the addition of constants factors; such constant
factors can be included within the definitions of $F_2$ and $F_4$
and eliminated by the previous transformation. In summary, the
only dependence on the coordinate $y$ of metric (\ref{eq:metric})
appears in the conformal factor $Q$.

Now, we apply the same strategy of Ref.~\cite{Garcia:2002gj},
namely we first redefine the functions $a$ and $b$ by
\begin{align}
a+b&=2k\,Y,& a&=k(Y+X),\nonumber\\
a-b&=2k\,X,& b&=k(Y-X). \label{eq:XY}
\end{align}
Using the new functions $X$ and $Y$, Eqs.~(\ref{eq:ax-bx}) and
(\ref{eq:ax*bx}) are rewritten as
\begin{eqnarray}
\frac{\mathrm{d}X}{\mathrm{d}x}&=&2kY^2\mathrm{e}^{-2P}, \label{eq:Xx}\\
\left(\frac{\mathrm{d}Y}{\mathrm{d}x}\right)^2
-\left(\frac{\mathrm{d}X}{\mathrm{d}x}\right)^2
&=&4\epsilon{k}^2Y^2\mathrm{e}^{-4P}. \label{eq:Yx}
\end{eqnarray}
Equation (\ref{eq:Xx}) suggests to choose a new coordinate $x$
defined by
\begin{equation}\label{eq:x2X}
\textstyle (\tau,\sigma,x,y)\mapsto
\left(\tau,\sigma,2k\int{Y^2\mathrm{e}^{-2P}\mathrm{d}x},y\right).
\end{equation}
The general solutions of Eqs.~(\ref{eq:Xx}) and (\ref{eq:Yx}) in
terms of the new coordinate $x$ are
\begin{eqnarray}
X(x)  &=&x,\\
Y^2(x)&=&(x-x_0)^2-\epsilon.
\end{eqnarray}

Up to now we have integrated seven equations; the four related
with $\Psi_0$ and $\Psi_4$, the imaginary part of $\Psi_2$, and
the real parts of $\Psi_3-\Psi_1$ and $\Psi_1$. Hence, it remains
to integrate the other three equations which allows to specify the
functions $P$, $F_1$, and $F_3$. These equations are the following
\begin{equation}\label{eq:Im(3-1)}
\Psi_3-\Psi_1=\frac{ik^{3/2}Y^{7/2}\mathrm{e}^{2Q-P}}{\sqrt{2}}
\frac{\mathrm{d}^2}{\mathrm{d}x^2}(F_3+kxF_1)=0,
\end{equation}
which integrates just as $F_3=-kxF_1$, since the two related
integration constants can be absorbed within the definitions of
the functions $F_1$ and $F_3$ and eliminated by a shifting of the
Killing coordinates as in the transformation (\ref{eq:shift}).
Also we use
\begin{equation}\label{eq:Re(2)}
\Psi_2=-\frac{k^2Y^3\mathrm{e}^{2Q}}{3}
\frac{\mathrm{d}^2}{\mathrm{d}x^2}\left(Y\mathrm{e}^{-2P}
+\frac{k}{2}Y^2{F_1}^2\right)=0,
\end{equation}
determining the function $P$ as
\begin{equation}\label{eq:P}
\mathrm{e}^{-2P}=\frac{C_0+C_1x}{Y}-\frac{k}{2}Y{F_1}^2.
\end{equation}
The last equation is
\begin{equation}\label{eq:Im(3+1)}
\Psi_3+\Psi_1=-\frac{ik^{5/2}Y^{5/2}\mathrm{e}^{2Q-P}}{\sqrt{2}}
\frac{\mathrm{d}^2}{\mathrm{d}x^2}(F_1Y^2)=0,
\end{equation}
giving
\begin{equation}
F_1(x)=\frac{K_0+K_1x}{(x-x_0)^2-\epsilon}.%\\
%F_3(x)&=&-\frac{kx(K_0+K_1x)}{(x-x_0)^2-\epsilon}.
\end{equation}

As last step, in order to write the obtained metric in simple form
we make the following coordinate transformation, see
Ref.~\cite{Garcia:2002gj},
\begin{equation}\label{eq:lastcoord}
\textstyle (\tau,\sigma,x,y)\mapsto
\left(\sqrt{2}(\tau+kx_0\sigma),\sqrt{2}k\sigma,x-x_0,2ky\right),
\end{equation}
together with the next redefinition of the conformal factor
$Q\mapsto{Q+\frac{1}{4}\ln(16k^4Y^2)}$ and also simple
redefinitions of the involved constants. The final form of the
most general conformally flat stationary cyclic symmetric metric
is
\begin{eqnarray}
\bm{g}&=&e^{-2Q(x,y)}\biggl(-k\left(\bm{d\tau}^2 +
2x\bm{d\tau}\bm{d\sigma} + \epsilon\bm{d\sigma}^2\right)\nonumber\\
&&{}+\frac{\bm{dx}^2}{(C_0+C_1x)(x^2-\epsilon)-k(K_0+K_1x)^2}\nonumber\\
&&{}+2k(K_0+K_1x)\bm{d\sigma}\bm{dy}+(C_0+C_1x)\bm{dy}^2\biggr).\,~
\label{eq:cfg}
\end{eqnarray}
It is easy to note that for $K_0=0=K_1$ we recover the circular
metrics of Ref.~\cite{Garcia:2002gj}. Instead, for $K_0\ne0$ and
$K_1\ne0$ the above metric is noncircular as follows from the
following quantities
\begin{eqnarray}
\nonumber *(\bm{m}\wedge\bm{k}\wedge\bm{dk})&=&
-k^2\mathrm{e}^{-2Q}(K_0+K_1x),\\
*(\bm{k}\wedge\bm{m}\wedge\bm{dm})&=&
-k^2\mathrm{e}^{-2Q}(\epsilon{K_1}+K_0x).
\end{eqnarray}

\section{\label{sec:staticity}Staticity}

As it was defined in Sec.~\ref{sec:SCSS}, the spacetimes derived
in the previous section would be static if there exist a timelike
linear combination of the Killing fields,
\begin{equation}\label{eq:staticcomb}
\bm{k}_\textrm{s}=A\bm{\frac{\partial}{\partial\tau}}
+B\bm{\frac{\partial}{\partial\sigma}},
\end{equation}
satisfying the staticity condition (\ref{eq:staticity}). For
metric (\ref{eq:cfg}) such condition becomes
\begin{eqnarray}
\nonumber 0=*(\bm{k}_\textrm{s}\wedge\bm{dk}_\textrm{s})&=&
k\bigl[\,B(K_0A-\epsilon{K_1}B)\bm{d\tau}\\
\nonumber
&&\quad{}+B(K_1A-K_0B)\bm{d\sigma}\\
&&\quad{}+(A^2-\epsilon{B}^2)\bm{dy}\bigr].
\end{eqnarray}
It is straightforward to realize that we are in the presence of
static configurations only if
\begin{equation}\label{eq:staticond}
\epsilon=1 \quad\mathrm{and}\quad K_1=\pm{K_0},
\end{equation}
in which case the hypersurface orthogonal Killing fields are
proportional to $\bm{k}_\textrm{s}=\bm{{\partial}/{\partial\tau}}
\pm\bm{{\partial}/{\partial\sigma}}$.

\section{\label{sec:axisymmetry}Axisymmetry}

Now we turn our attention to the existence of a rotation axis,
i.e., there are conformally flat stationary axisymmetric
configurations within the class (\ref{eq:cfg}). It follows from
the definition of Sec.~\ref{sec:SCSS} that the rotation axis is
the spacetime region where the cyclic Killing field $\bm{m}$
vanishes. For metric (\ref{eq:cfg}) its general stationary and
cyclic Killing fields are a linear combination of the vectors
$\bm{\partial_\tau}$ and $\bm{\partial_\sigma}$. Hence, performing
the transformation $(\tau=\alpha t+\beta \phi,\sigma=\gamma
t+\delta \phi)$, where $\alpha\delta-\beta\gamma\neq0$, the
Killing fields are written as $\bm{k}=\bm{\partial_t}$ and
$\bm{m}=\bm{\partial_\phi}$, respectively. In terms of the new
coordinates, all the metric components
$g_{\phi\mu}=\bm{g}(\bm{m},\bm{\partial_\mu})$ must vanish on the
axis, which implies the following set of algebraic equations
\begin{subequations}
\begin{eqnarray}\label{eq:axis_set}
g_{\phi{t}}&=&-ke^{-2Q}[(\alpha\delta+\beta\gamma)x
+\alpha\beta+\epsilon\gamma\delta]=0,\label{eq:axis_pt}\\
g_{\phi\phi}&=&-ke^{-2Q}(\beta^2+2\beta\delta{x}+\epsilon\delta^2)=0,
\label{eq:axis_pp}\\
g_{\phi{y}}&=&ke^{-2Q}\delta(K_0+K_1x)=0.\label{eq:axis_py}
\end{eqnarray}
\end{subequations}
Isolating $x$ from Eq.~(\ref{eq:axis_pp}) and inserting the result
in Eq.~(\ref{eq:axis_pt}) we obtain
\begin{equation}\label{eq:b2d}
\frac{(\alpha\delta-\beta\gamma)(\beta^2-\epsilon\delta^2)}
{\beta\delta}=0,
\end{equation}
since $\alpha\delta-\beta\gamma\neq0$ the above equation has
nontrivial solutions only if $\epsilon=1$. Using the above
conditions in the remaining Eq.~(\ref{eq:axis_py}) we obtain
\begin{equation}\label{eq:K12K0}
\delta(K_1\mp{K_0})=0,
\end{equation}
which implies that the related rotation axis must be located at
\begin{equation}\label{eq:axis}
x=\mp1.
\end{equation}

Summarizing, metric (\ref{eq:cfg}) describes a stationary
axisymmetric spacetime only for $\epsilon=1$ and $K_1=\pm{K_0}$,
i.e., the conditions for the existence of the rotation axis are the
same than guaranty than the spacetime is static, see conditions
(\ref{eq:staticond}).

As a consequence, the known Collinson theorem
\cite{Collinson:1976,Garcia:2002gj} not only is generalized to
include configurations which are not necessarily circular, but, it
is part of a more general statement: any conformally flat stationary
cyclic symmetric spacetime, even a noncircular one, is additionally
axisymmetric if and only if it is also static.

\section{\label{sec:conclu} Conclusions}

In this paper we study all the stationary cyclic symmetric
spacetimes which are at the same time conformally flat. In contrast
to previous work on the subject we consider also noncircular
configurations. The conformal flatness is imposed by demanding the
vanishing of the Weyl tensor. The resulting constraints are
extremely involved by the inclusion of noncircular contributions,
and leave us with a system of ten nonlinear pdes, see Appendix
\ref{app:Weyl}. However, it is still possible to achieve its full
integration as we show in detail at Sec.~\ref{sec:CTh}. The class of
obtained spacetimes is fully determined up to a conformal factor
which respects the spacetime symmetries, and several integration
constants. In particular, two integration constants characterize the
noncircular behavior of these spacetimes, when they vanish we
recover the circular configurations obtained in
Ref.~\cite{Garcia:2002gj} by two of the authors.

We investigate the conditions allowing the existence of a rotation
axis in the resulting configurations. The static spacetimes within
the class are also considered. It results that the parameters
election for these two physical situations is the same: $\epsilon=1$
and $K_1=\pm{K_0}$, i.e., the involved spacetimes are axisymmetric
if and only if they are also static. Hence, one of the main result
of the paper can be summarized in the following
\begin{quote}
\textbf{Theorem}: \emph{Any conformally flat stationary cyclic
symmetric spacetime, even a noncircular one, is additionally
axisymmetric if and only if it is also static.}
\end{quote}
With regard to the properly cyclic symmetric class (with no rotation
axis, and by the above theorem containing necessarily nonstatic
spacetimes) it will be interesting to investigate what kind of
sources can solve Einstein equations with gravitational fields
within this class. In the case that be possible to retain the
noncircular contributions in this process, the derived
configurations would corresponds to the first exact noncircular
gravitational fields found in the literature.

\begin{acknowledgments}
We thank M.~Hassa\"{\i}ne for discussions. This work has been
partially supported by FONDECYT Grants 1040921, 7040190, and
1051064, CONACyT Grants 38495E and 34222E, CONICYT/CONACyT Grant
2001-5-02-159, and MECESUP Grant FSM 0204. The generous support of
Empresas CMPC to the Centro de Estudios Cient\'{\i}ficos (CECS) is
also acknowledged. CECS is a Millennium Science Institute and is
funded in part by grants from Fundaci\'on Andes and the Tinker
Foundation.
\end{acknowledgments}

\appendix

\section{\label{app:Weyl}Weyl Complex Components}

For the null tetrad (\ref{eq:terad}) the complex components of the
Weyl tensor are given by
%\begin{widetext}
\begin{eqnarray}
\Psi_0&=&\frac{2\mathrm{e}^{2(Q+P)}}{a+b}
\left[\frac{\partial^2a}{\partial{z}^2}
+2\frac{\partial{P}}{\partial{z}}\frac{\partial{a}}{\partial{z}}
-\frac{2}{a+b}\left(\frac{\partial{a}}{\partial{z}}\right)^2
\right],\\
\bar{\Psi}_4&=&\frac{2\mathrm{e}^{2(Q+P)}}{a+b}
\left[\frac{\partial^2b}{\partial{z}^2}
+2\frac{\partial{P}}{\partial{z}}\frac{\partial{b}}{\partial{z}}
-\frac{2}{a+b}\left(\frac{\partial{b}}{\partial{z}}\right)^2
\right],\,\qquad\\
6\Psi_2&=&\frac{2\mathrm{e}^{2(Q+P)}}{(a+b)^2}\left(2(a+b)^2
\frac{\partial^2P}{\partial{z}\partial\bar{z}}+5\frac{\partial{a}}{\partial{z}}
\frac{\partial{b}}{\partial\bar{z}}-\frac{\partial{a}}{\partial\bar{z}}
\frac{\partial{b}}{\partial{z}}\right)\nonumber\\
&&{} -\frac{4\mathrm{e}^{2Q+4P}}{a+b}\mathrm{Re}\left(
\frac{\partial{M}}{\partial\bar{z}}-\frac{M+N}{a+b}
\frac{\partial{a}}{\partial\bar{z}}\right)\nonumber\\
&&{}\qquad\qquad\times\mathrm{Re}\left(
\frac{\partial{N}}{\partial\bar{z}}-\frac{M+N}{a+b}
\frac{\partial{b}}{\partial\bar{z}}\right),\\
%\end{eqnarray}
%\begin{eqnarray}
2\Psi_1&=&\frac{\mathrm{e}^{2Q+P}}{i\sqrt{a+b}}\frac{\partial}{\partial{z}}
\left[\mathrm{e}^{2P}\mathrm{Re}\left(
\frac{\partial{M}}{\partial\bar{z}}-\frac{M+N}{a+b}
\frac{\partial{a}}{\partial\bar{z}}\right)\right]\nonumber\\
&&{}-\frac{\mathrm{e}^{2Q+3P}}{i(a+b)^{3/2}}\frac{\partial{a}}{\partial{z}}
\nonumber\\
&&{}\times\mathrm{Re}\biggl[\frac{\partial{M}}{\partial\bar{z}}
-3\frac{\partial{N}}{\partial\bar{z}}
-\frac{M+N}{a+b}\left(\frac{\partial{a}}{\partial\bar{z}}
-3\frac{\partial{b}}{\partial\bar{z}}\right)\biggr],\,\qquad\\
2\bar{\Psi}_3&=&\frac{i\mathrm{e}^{2Q+P}}{\sqrt{a+b}}\frac{\partial}{\partial{z}}
\left[\mathrm{e}^{2P}\mathrm{Re}\left(
\frac{\partial{N}}{\partial\bar{z}}-\frac{M+N}{a+b}
\frac{\partial{b}}{\partial\bar{z}}\right)\right]\nonumber\\
&&{}-\frac{i\mathrm{e}^{2Q+3P}}{(a+b)^{3/2}}\frac{\partial{b}}{\partial{z}}
\nonumber\\
&&{}\times\mathrm{Re}\biggl[\frac{\partial{N}}{\partial\bar{z}}
-3\frac{\partial{M}}{\partial\bar{z}}-\frac{M+N}{a+b}\left(
\frac{\partial{b}}{\partial\bar{z}}
-3\frac{\partial{a}}{\partial\bar{z}}\right)\biggr].\,\qquad
\end{eqnarray}
%\end{widetext}

\section{\label{app:a=b}The case $a=b$}

In section \ref{sec:CTh} it was shown that $a-b=F(x)$ and the
complete study of the case $\mathrm{d}F/\mathrm{d}x\neq0$ was
performed. Here, we concentrate in the special case
$F=\mathrm{const.}$, such constant can be putted to zero by
shifting the timelike coordinate and redefining function $a$ or
function $b$. Hence, this case is equivalent to have $a=b$. Since
now $\Psi_0=\bar{\Psi}_4$, equation (\ref{eq:0-4}) [or its
consequence (\ref{eq:Xx})] is pointless, which invalidates the
coordinate transformation (\ref{eq:x2X}). This is the reason why
this case must be studied separately. The vanishing of the
$\Psi_0$ component for $a=b$ implies
\begin{equation}\label{eq:0a}
\Psi_0=\mathrm{e}^{2Q}\frac{\partial}{\partial{z}}
\left(\mathrm{e}^{2P}\frac{\partial}{\partial{z}}\ln{a}\right)=0\Rightarrow
\frac{\partial{a}}{\partial{z}}=\bar{g}(\bar{z})a\,\mathrm{e}^{-2P}.
\end{equation}
We can take $g(z)=1$ again, just applying the coordinate
transformation (\ref{eq:rescal}) together with the relevant
redefinitions of the functions $P$, $M$, and $N$. Applying the
same arguments than in section \ref{sec:CTh} we conclude that $a$
and $P$ are functions of coordinate $x$ only. Choosing now the
gauge elections $\bar{M}=M$ and $\bar{N}=N$ we end with metric
(\ref{eq:metricy}) evaluated in $b=a$. Equations
(\ref{eq:Re(3-1)}) and (\ref{eq:Re(3)}) and their solutions
(\ref{eq:M}) and (\ref{eq:N}) are the same just considering that
$b=a$.

Since $a$ and $P$ are functions of $x$ only, the first order
equation (\ref{eq:0a}) suggests to take $a$ as a new spatial
coordinate by
\begin{equation}\label{eq:x2a}
\textstyle (\tau,\sigma,x,y)\mapsto
\left(\tau,\sigma,a=\mathrm{exp}({\int{\mathrm{e}^{-2P}\mathrm{d}x}}),y\right).
\end{equation}
In terms of this coordinate Eqs.~(\ref{eq:Im(3-1)}) and
(\ref{eq:Im(3+1)}) are written in this case as
\begin{eqnarray}\label{eq:Im(3-1)a}
\Psi_3-\Psi_1&=&\frac{ia^{5/2}\mathrm{e}^{2Q-P}}{4\sqrt{2}}
\frac{\mathrm{d}}{\mathrm{d}a}
\left(\frac{1}{a}\frac{\mathrm{d}F_3}{\mathrm{d}a}\right)=0,\\
\label{eq:Im(3+1)a}
\Psi_3+\Psi_1&=&-\frac{i\mathrm{e}^{2Q-P}}{4\sqrt{2}\sqrt{a}}
\frac{\mathrm{d}}{\mathrm{d}a}
\left(a^3\frac{\mathrm{d}F_1}{\mathrm{d}a}\right)=0,
\end{eqnarray}
giving
\begin{equation}
F_1(a)=\frac{K_0}{a^2},\qquad \label{eq:F1aF3a} F_3(a)=K_1a^2.
\end{equation}
Using these expressions, Eq.~(\ref{eq:Re(2)}) becomes
\begin{equation}\label{eq:Re(2)a}
\Psi_2=-\frac{a^2\mathrm{e}^{2Q}}{12}
\frac{\mathrm{d}}{\mathrm{d}a}\left[\frac{1}{a}\frac{\mathrm{d}}{\mathrm{d}a}
\left(a\mathrm{e}^{-2P}
+\frac{{K_0}^2}{2a^2}-\frac{{K_1}^2a^4}{2}\right)\right]=0,
\end{equation}
and integrates as
\begin{equation}\label{eq:Pa}
\mathrm{e}^{-2P(a)}=\frac{\alpha+\beta{a}^2}{a}
-\frac{{K_0}^2}{2a^3}+\frac{{K_1}^2a^3}{2}.
\end{equation}
Finally, for a general stationary cyclic symmetric spacetime with
$a=b$, the conformally flat metrics are given by
\begin{eqnarray}
\bm{g}&=&e^{-2Q(a,y)}\biggl[-\frac{\bm{d\tau}^2}{a}
+a\bm{d\sigma}^2\nonumber\\
&&{} +\frac{\bm{da}^2}{a(\alpha+\beta{a}^2)
-\frac{{K_0}^2}{a}+{K_1}^2a^5}\nonumber\\
&&{}+2\left(\frac{K_0}{a}\bm{d\sigma}-K_1a\bm{d\tau}\right)\bm{dy}
+\frac{\alpha+\beta{a}^2}{a}\bm{dy}^2\biggr].\,\quad
\label{eq:cfga}
\end{eqnarray}
where we have rescaled the Killing coordinates and some of the
constants. For $K_0=0=K_1$ we recover the static metric of
Ref.~\cite{Garcia:2002gj}. In general, the above metric describes a
static spacetime for $K_0=0$. Additionally, it is incompatible with
the existence of a rotation axis, hence, this class is not in
contradiction with the Collinson theorem.


\begin{thebibliography}{9}

\bibitem{Gourgoulhon:1993}
  E.~Gourgoulhon and S.~Bonazzola,
  Phys. Rev. D \textbf{48}, 2635 (1993).

\bibitem{Ioka:2003dd}
  K.~Ioka and M.~Sasaki,
  %``Grad-Shafranov equation in noncircular stationary axisymmetric
  %spacetimes,''
  Phys.\ Rev.\ D \textbf{67}, 124026 (2003)
  [arXiv:gr-qc/0302106].
  %%CITATION = GR-QC 0302106;%%

\bibitem{Ioka:2003nh}
  K.~Ioka and M.~Sasaki,
  %``Relativistic stars with poloidal and toroidal magnetic fields and
  %meridional flow,''
  Astrophys.\ J.\ \textbf{600}, 296 (2004)
  [arXiv:astro-ph/0305352].
  %%CITATION = ASTRO-PH 0305352;%%

\bibitem{Vera:2003cn}
  R.~Vera,
  %``Influence of general convective motions on the exterior of isolated
  %rotating bodies in equilibrium,''
  Class.\ Quant.\ Grav.\ \textbf{20}, 2785 (2003)
  [arXiv:gr-qc/0305108].
  %%CITATION = GR-QC 0305108;%%

\bibitem{Kundt:1966}
  W.~Kundt and M. Tr\"{u}mper,
  Z. Phys. \textbf{192}, 419 (1966).

\bibitem{Collinson:1976}
  C.~D.~Collinson,
  Gen. Rel. Grav. \textbf{7}, 419 (1976).

\bibitem{Garcia:2002gj}
  A.~A.~Garc\'{\i}a and C.~Campuzano,
  %``On conformally flat stationary axisymmetric spacetimes,''
  Phys.\ Rev.\ D \textbf{66}, 124018 (2002)
  [Erratum-ibid.\ D \textbf{68}, 049901 (2003)]
  [arXiv:gr-qc/0205080].
  %%CITATION = GR-QC 0205080;%%

\bibitem{Barnes:2003gz}
  A.~Barnes and J.~M.~M.~Senovilla,
  %``Comment on Conformally flat stationary axisymmetric metrics,''
  arXiv:gr-qc/0305091.
  %%CITATION = GR-QC 0305091;%%

\bibitem{Garcia:2003is}
  A.~A.~Garc\'{\i}a and C.~Campuzano,
  %``Circular stationary cyclic symmetric spacetimes: conformal flatness,''
  arXiv:gr-qc/0310054.
  %%CITATION = GR-QC 0310054;%%

\bibitem{Heusler:1996}
  M.~Heusler,
  \emph{Black Hole Uniqueness Theorems}
  (Cambridge Univ. Press, Cambridge 1996).

\end{thebibliography}
\end{document}